\documentclass[namedreferences,hyperref,optionalrh]{spr-sola}
\usepackage{graphicx}        
\usepackage{color}           

\chardef\us=`\_
\begin{document}

\begin{frontmatter}


\title{Coronal Signatures of Flare Generated Fast Mode Wave at EUV and Radio wavelengths\\ {\it Solar Physics}}

\author[addressref={aff1},email={vasanth.veluchamy@uj.edu.pl}]{V.~\surname{Vasanth}\orcid{0000-0002-6056-7899} \sep}

       \address[id=aff1]{Astronomical Observatory of Jagiellonian University, Krakow 30-244, Poland}

\runningauthor{Vasanth}
\runningtitle{Coronal Signatures of Flare Generated Wave at EUV and Radio wavelength}

\begin{abstract}
 This paper presents a detailed study of the type II solar radio burst that occurred on 06 March 2014 using combined data analysis. It is a classical radio event consisting of type III radio burst and a following type II radio burst in the dynamic spectrum. The type II radio burst is observed between 235 $\--$ 130 MHz (120 $\--$ 60 MHz) in harmonic (fundamental) bands with the life time of 5 minutes between 09:26 UT $\--$ 09:31 UT. The estimated speed of type II burst by applying twofold Saito model is $\sim$ 650 km s$^{-1}$. An \textrm{extreme ultraviolet (EUV)} wave is observed with \textrm{Atmospheric Imaging Assembly (AIA)} onboard the \textrm{Solar Dynamics Observatory (SDO)} pass bands. The very close temporal onset association of \textrm{EUV} wave and flare energy release indicates that the \textrm{EUV} wave is likely produced by a flare pressure pulse. The eruption is also accompanied by a weak coronal mass ejection (CME) observed with the coronagraphs onboard \textrm{Solar and Heliospheric Observatory (SOHO)} and \textrm{the twin Solar Terrestrial Relations Observatory (STEREO)}. The plane of sky speed of the CME was $\sim$ 252 km s$^{-1}$ atSOHO/LASCO-C2 and $\sim$ 280 km s$^{-1}$ at STEREO-B/SECCHI-COR1 FOV. The EUV wave has two wave fronts, one expanding radially outward, and the other one moving along the arcade. The source position of the type II burst imaged by the \textrm{Nan{\c{c}}ay Radio Heliograph (NRH)} shows that it was associated with the outward moving \textrm{EUV} wave. The CME is independent of the shock wave as confirmed by the location of \textrm{NRH} radio sources below the CME's leading edge. Therefore the type II radio burst is probably ignited by the flare. This study shows the possibility of \textrm{EUV} wave and coronal shock triggered by flare pressure pulse, generating the observed type II radio burst.

\end{abstract}

\end{frontmatter}

\section{Introduction}
     \label{S-Introduction}

The coronal disturbances and shock waves are formed during the energy release processes such as flares and coronal mass ejections (CMEs). Both CMEs and solar flares are efficient in accelerating electrons as well as in producing shock waves whose radio signatures in the dynamic spectra are called solar type II radio bursts \citep{McLean85, Dulk85, Reames99,Vrsnak08}. Type II radio bursts are the earliest known signature of shock waves, drifting from high to low frequency in the dynamic spectrum due to emission at the local plasma frequency and/or its harmonics generated via plasma emission mechanism \citep{Payne-Scott47, Wild50, Nelson85}. The coronal type II burst generally occurs at the frequency range between 300 $\--$ 20 MHz corresponding to the height of 1.3 $\--$ 3 $R_\odot$. The interplanetary type II bursts at decametric, hectometric and kilometric wavelengths occur at the frequency range between 14 MHz $\--$ 20 kHz generally beyond 2 $R_\odot$ and are known to be driven by fast and wide CMEs \citep{Gopalswamy01, Lara03, Vasanth13, Vasanth13b, Prakash12, Shanmugaraju18}. IP type II bursts recorded at large heliocentric distances can be considered as a possible remote sensing indicator for forecasting space weather effects \citep{Gopalswamy01, Vasanth15}.

Solar flares, CMEs and shock waves are closely associated with one another and their association rate increases with the flare importance \citep{Vrsnak08}. Also there is a good synchronization between CME acceleration and impulsive phase of flares \citep{Zhang01, Zhang04, Vrsnak04, Maricic07, Temmer10}. Due to this it is often difficult to identify the origin of coronal shocks, i.e. to distinguish whether they are ignited by a flare or initiated by a CME. In recent years, a number of case studies have shown that coronal shocks are driven by CMEs. However, there are cases of type II bursts observed in absence of CMEs or associated with slow CMEs which indicate that they might be generated by flare blast waves  \citep{Vrsnak00, Vrsnak01, Vrsnak08, Magdalenic08, Magdalenic10, Magdalenic12, Vasanth11, Vasanth14, Kumar13, Kumar16}.

Solar eruptions are frequently accompanied by large-scale coronal wave phenomena known as \textrm{EUV wave} first observed by the \textrm{Extreme-ultraviolet Imaging Telescope (EIT)} onboard the \textrm{SOHO} spacecraft \citep{Thompson99}. Some reports using \textrm{SOHO/EIT} images claim that EUV waves are related to pressure pulse of a flare \citep{Wu01, Vrsnak02, Vrsnak08}. The shock waves of such events are generally associated with impulsive flares of short duration. On the other hand, a number of studies argue that weak flares cannot induce large-scale waves, which indicates that EUV waves are caused by CMEs. Even now the source of \textrm{EUV wave} is under debate \citep{Long17}.

The previous studies had reported that the CMEs propagating with speed $<$ 400 km s$^{-1}$ are not likely to produce the coronal shocks. However, \cite{Pohjolainen06} found evidence that flare related CME acceleration can excite type II bursts even if the CME remains slow. Apparently, local conditions have to be favourable for the shock formation, one of which is certainly low Alfv{\'e}nic speed.

This paper presents a case study of the metric type II burst observed on 06 March 2014. This event was previously analysed by \cite{Kumar15} using \textrm{SDO/AIA} \textrm{field of view (FOV)} and concluded that a fast-mode EUV wave was generated by flare pressure pulse, which excited metric type II burst. The present paper will focus on the context of radio observations and their relations to solar eruptions. The fact is that a CME was recorded in STEREO$/$SECCHI and SOHO/LASCO FOV, not discussed in the previous study. Therefore, the relationship between the type II burst, EUV wave and CME structure will be explored in detail using \textrm{NRH} observations.\\

\section{Observations}
    \label{S-Observations}

    The dynamic spectrum of the type II radio burst was recorded on 06 March 2014 by the \textrm{ORFEES
radio spectrograph} from 500 $\--$ 144 MHz and the \textrm{Learmonth} from 180 $\--$ 25 MHz. The composite of the spectral data is presented in Figure 1a.
The burst showed a fundamental and harmonic emission, both are well observed with
duration of 5 minutes between 09:26 UT $\--$ 09:31 UT. Two minutes before the type II burst, a type III radio
burst was also recorded and shown in Figure 1a. The type II appears only in the metric
domain \textrm{i.e.,} no longer wavelength type II radio burst was recorded by \textrm{Wind/WAVES} spectrum \citep{Bougeret95} and either by \textrm{STEREO-B/WAVES} \citep{Bougeret08}.

This study uses radio imaging from the \textrm{NRH} at selected frequencies 408, 327, 298, 270, 173, and 150 MHz \citep{Kerdraon97}. The spatial resolution of \textrm{NRH} images depends on the imaging frequency and observational
time, being 1$'$ - 3$'$ at 327 MHz, decreasing to 3$'$ - 6$'$ at 150 MHz. The highest temporal resolution
of \textrm{NRH} is 0.25 s. The present study utilized the data with 0.25 s resolution and integrated them to 1 s.

The type II radio burst is associated with a solar eruption from the NOAA active region (AR11998)
at the south-eastern limb. The eruption is accompanied by a compact or small flare located around S07E64,
which starts at 09:23 UT, peaks at 09:25 UT and ends at 09:30 UT according to the \textrm{Reuven Ramaty High
Energy Solar Spectroscopic Imager (RHESSI)} hard X-ray data \citep{Lin02} (see Figure 1). There is no GOES soft X-ray flare observation available during the time. The accompanying CME is observed by several instruments at various \textrm{EUV} wavelengths from the inner to the outer corona and with coronagraphs in white
light (WL). These instruments include \textrm{AIA} \citep{Lemen12} onboard \textrm{SDO}
that records the full disk image of the Sun with a cadence of 12 s \citep{Pesnell12} and the EUV Imagers \textrm{(EUVI)}
onboard \textrm{STEREO/SECCHI} with one ahead and the other behind the Earth separated  by $\sim$ 46 degrees \citep{Kaiser08}, The longitudinal position of STEREO-A and B with respect to the Earth are -161 and 153 degrees, making a separation of $\sim$ 46 degrees. Both the A and B were near the far side of the Sun. The observed eruption is a south-eastern limb event in the Sun
\textrm{[SDO/AIA and SOHO/LASCO FOV]}, and is observed only in \textrm{STEREO-B} and not by \textrm{STEREO-A}.  At larger heights, the CME is observed by the \textrm{Large Angle Spectrometric Coronagraph (LASCO) C2} \citep{Brueckner95} onboard the \textrm{SOHO} spacecraft \citep{Domingo95}. The observed CME structure is
diffuse in \textrm{SDO/AIA FOV}, but is clear and well observed in \textrm{STEREO-B/SECCHI-COR1 and SOHO/LASCO-C2 images (Figure 3)}. The speed of the CMEs was estimated as $\sim$ 252 km s$^{-1}$ at \textrm{SOHO/LASCO-C2} and $\sim$ 280 km s$^{-1}$ at \textrm{STEREO-B/SECCHI-COR1} FOV.

\section{Data-Analysis and Results}
    \subsection{Solar radio emission and EUV-white light eruptive structures}

The type II solar radio burst in the dynamic spectrum  [Figure 1a], shows the drifting feature in both the fundamental (120 $\--$ 60 MHz) and harmonic bands (235 $\--$ 135 MHz). It is well established by the earlier reports that in meter wavelength radio emission the harmonic band is more intense and better defined than the fundamental one \citep{Vrsnak01}. Using $2\times$Saito electron density model \citep{Saito70}, the height of the type II burst is estimated by selecting several points from the harmonic band, then they are converted to fundamental for further estimations. The estimated heights are shown in the height$\--$time plot in Figure 6. The estimated speed of type II burst is $\sim$ 650 km s$^{-1}$.

The type II burst is found to be accompanied by a compact flare located at S07E64 by \textrm{RHESSI} observations (Figure 1). It further confirms the presence of non-thermal electrons at flare timing. There is a weak CME associated with this eruption but not reported in the previous study by \cite{Kumar15}.

The \textrm{EUV} images showed a fast-propagating wave, probably the shock wave, that moved outward and could have generated the type II burst. The \textrm{EUV} wave is very clearly observed in the \textrm{AIA 171 {\AA}} channel and the base difference images were used to determine the kinematics and propagation speed using slice-cut between 09:23 UT $\--$ 09:26 UT. The estimated mean speed of the wave is $\sim$ 820 km s$^{-1}$, obtained by linear fit to the selected data points in the interval. Both the flare energy release and onset of \textrm{EUV} wave occurred simultaneously around 09:23 UT (see \cite{Kumar15}, for details). Therefore the wave is most likely generated by the flare pressure pulse.  The \textrm{EUV} wave at \textrm{AIA 171 {\AA}} has two components, one propagating perpendicular and another along (parallel to) the flare region. While the parallel propagating component gets reflected backward from the other leg of the flare loops (Figure 2a and 2c), the perpendicular component freely propagates outward. After 09:32 UT, a rising filament can been seen above the active region. The EUV wave is also observed in \textrm{AIA 211 {\AA} and AIA 193 {\AA}. In the AIA 211 {\AA}} images it is observed as a freely propagating outward moving structure (Figure 2d).

After checking the coronagraph images, it was found that the eruption is accompanied by a CME. The evolution of the CME observed by \textrm{STEREO-B/SECCHI-EUVI}, \textrm{STEREO-B/SECCHI-COR1 and SOHO/LASCO} is shown in Figure 3. The earliest signature and evolution of the CME revealed in EUV wavelength by \textrm{STEREO-B/SECCHI-EUVI 195 {\AA}} running difference images at 09:26 UT, 09:31 UT and 09:36 UT are presented in Figure 3b-d, showing a slow evolution of the loop system that started at the impulsive phase of the flare.

The CME is observed at \textrm{STEREO-B/SECCHI-EUVI} at the same time when the \textrm{EUV wave} is at the edge of \textrm{AIA FOV}. The \textrm{AIA} instruments and the twin \textrm{STEREO} spacecraft observations were compared and found that the CME loop structure is moving ahead and independent of the observed {\textrm{EUV} wave. Further in \textrm{STEREO-B/SECCHI-COR1 FOV} the expanding loop structure is visible at 09:36 UT. The expanding structure, probably the CME, will take several minutes to reach the \textrm{SOHO/LASCO-C2 FOV} at heights $>$ 2.2 $R_\odot$. Therefore, the structure is visible at \textrm{SOHO/LASCO-C2 FOV} at 10:12 UT. This observation clearly shows that the eruption is accompanied by a CME and previous authors \citep{Kumar15} had missed it, likely due to not including \textrm{STEREO/SECCHI} observations in their analysis. They claim that there was no CME loop behind the \textrm{EUV} wave.

\subsection{Radio Imaging}

The \textrm{NRH} provides the radio imaging at nine different frequencies from 450 $\--$ 150 MHz. The highly sensitive \textrm{NRH} imaging observations were used to find the temporal position and outward motion of radio sources from the eruptive site. They can also be utilized to identify the onset of radio emissions or high frequency counterpart of the radio emissions that are possibly not recorded in the dynamic spectrum due to low instrument resolution. The present study utilized the \textrm{NRH} imaging data with temporal resolution of 0.25 s. The total radio flux profiles recorded at 270, 228, 173 and 150 MHz were plotted over the dynamic spectrum in Figure 1a as colored curves to identify the onset of the type II burst. The evolution of the dominant source is compared with radio features in the radio dynamic spectrum. The radio sources are identifiable only for the harmonic band due to the instrumental limit ${\textrm i.e.}$ the last observable NRH radio imaging observations. Using the Nan{\c{c}}ay radio imaging data, the different kinds of radio bursts can be easily identified by their shift of the total flux maximum at subsequent frequencies with time.
The total flux data observed by \textrm{NRH} single frequencies are consistent with the type III and type II bursts in the dynamic spectrum.

Figure 4 shows the temporal evolution of radio sources at different \textrm{NRH} frequencies: 228, 173 and 150 MHz.
An accompanying animation is provided as electronic supplementary material to show the temporal and spatial evolution
of the emissions at the six \textrm{NRH} frequencies between 327 $\--$ 150 MHz. The contours represent the 95 and 90\% levels of the $T_{\rm B max}$.
In the animation, we see that before the type III burst, the AR released weak radio emissions with a $T_{\rm B max}$ of 10$^{6}$ K. The type III radio burst start at 09:23:05 UT, attains the $T_{\rm B max}$ of 10$^{10}$ K at 09:23:20 UT and ends at 09:24:10 UT.

The type II burst first appeared in the \textrm{NRH FOV} at 09:25:50 UT at 228 MHz. Later
on, it appeared at successively lower frequencies. At 173 MHz, the $T_{\rm B max}$ is
4.5 x 10$^{10}$ K  at 09:27:39 UT, whereas it is 10$^{9}$ K
at 09:27:05 UT at 228 MHz, and also at 09:29:07 UT at 150 MHz. The $T_{\rm B max}$ reaches higher values at lower frequencies. The type II sources can be identified from 228 MHz to lower frequency channels one after the other with some delay consistent with the drifting motion from higher to lower frequencies observed in the radio dynamic spectrum. The radio sources disappearing at high altitude indicates either the type II emission had ceased or moved out of \textrm{NRH FOV}.

In Figure 5, the \textrm{NRH} sources from 408 $\--$  150 MHz were plotted over the temporally closest \textrm{SDO/AIA 171 {\AA}} images to examine the temporal and spatial evolution of radio emission relative to the eruptive structures. The line-up of type III radio sources from higher to lower frequencies simultaneously at all frequency channels implies the propagation of electron beam along open field or large scale loop structure (Ginzburg and Zhelezniakov 1958; Melrose 1980). Energetic electrons are usually believed to be accelerated by magnetic reconnection in solar flares. In the case of type II radio burst, the observed \textrm{NRH} data shows the shift of radio sources from higher to lower frequencies with time. The type II radio sources at 173 and 150 MHz, located at the flank of the \textrm{EUV} wave observed at \textrm{AIA 211~\,\AA{}} channel presented by dotted blue line in Figure 5d-f, indicates the presence of a shock. In the \textrm{STEREO-B/SECCHI-EUVI} image the radio sources are located at lower heights than the frontal eruptive structure. This height difference between the CME leading edge and the shock wave implies that the type II burst is not driven by the CME bow shock.

 As the NRH has imaged the type II burst only in the second harmonic emission. It is important to note that the radio sources corresponding to harmonic and fundamental bands will appear close to each other. Yet their sources may look different due to refraction and scattering. The radio source corresponding to fundamental band might be twice larger in size than their harmonic counterpart. The shift in position of the radio source due to ionospheric refraction is $<$ 0.1 $R_\odot$ at 160 MHz. Therefore, the refraction effects will be minimal at higher frequencies (for details see Stewart and McClean, 1982). The error in estimating the type II burst location is 0.08 $\--$ 0.2 $R_\odot$ and is due to angular resolution of the NRH observations.

In the case of piston driven shock, the shock wave will be propagating ahead of the CME. If the situation is such that the shock wave is produced at the CME flank and not on the CME front, then the height of the shock structure will be smaller than the height of CME front. So, the possibility of shock being on the CME flank is unlikely.  Furthermore, it is important to note that for the case of CME-driven shock (bow shock), the CME speed would be close to the speed of shock wave. In contrast, the CME have a speed of $\sim$ 280 km s$^{-1}$ and type II speed is $\sim$ 650 km s$^{-1}$, which implies that the CME was probably not driving the shock wave. Thus it is concluded that CME might not be the driver of shock wave in the present case.

\subsection{Kinematic Evolution}

Figure 6 show the kinematics of radio emission and the eruptive structure observed between 09:20 UT $\--$ 11:20 UT across multitude wavelength radio, \textrm{EUV} and white-light observations from different instruments. The height$\--$time plots from \textrm{STEREO-B/SECCHI-EUVI, STEREO-B/SECCHI-COR1 and SOHO/LASCO-C2 FOV} show that the same feature is continuously tracked across different instruments. The EUV wave from \textrm{SDO/AIA at 211 {\AA}} observed between 09:24 $\--$ 09:27 UT is located at 1.003 $\--$ 1.22 $R_\odot$ and the radio sources at different NRH frequencies, i.e. 228 MHz radio sources located between 1.16 $R_\odot$ at 09:26 UT to 1.19 $R_\odot$ at 09:27 UT, 173 MHz radio sources located between 1.20 $R_\odot$ at 09:27 UT to 1.25 $R_\odot$ at 09:29 UT and 150 MHz radio sources located between 1.25 $R_\odot$ at 09:28 UT to 1.32 $R_\odot$ at 09:30 UT respectively. Therefore, the \textrm{EUV} wave from \textrm{SDO/AIA at 211 {\AA}} and type II radio burst imaged by NRH  were synchronized indicating the flare shock as the possible factor for the origin of radio emission. The measured speed of CME was in the range of 252 $\--$ 280 km s$^{-1}$, and such slow CME might not be able to drive a shock. On the other hand, the flare-triggered fast mode wave could lead to the shock wave as it propagates at a speed of $\sim$ 820 km s$^{-1}$. The radio measurements from the dynamic spectra and employing $2\times$Saito density model corresponds to the height of 1.18 $R_\odot$ at 09:26 UT to 1.35 $R_\odot$ at 09:29 UT, further the h$\--$t measurements yield a speed of $\sim$ 650 km s$^{-1}$. This also indicates that the type II radio emission in the present case might be excited by the flare shock and not by the CME.

The rising loop-like feature observed by \textrm{STEREO-B/SECCHI-EUVI} is the earliest possible CME leading-edge observation detected between 1.5 $\--$ 1.71 $R_\odot$ and thereafter continuously observed in \textrm{STEREO-B/SECCHI-COR1} between 1.58 $\--$ 2.44 $R_\odot$ and later in \textrm{SOHO/LASCO-C2 FOV} from 2.35 $\--$ 4.15 $R_\odot$. This shows that the high-located EUV loops started to move outward and the EUV wave from behind created the type II burst and probably made the lower-located CME materials to move faster. The impulsive flare triggered the fast-moving \textrm{EUV} wave. The height$\--$time measurements from the dynamic spectrum and, the type II sources imaged by NRH observations at 228, 173, 150 MHz synchronized with each other indicates the flare shock as the possible factor for the origin of type II radio burst. Also cannot rule out the possibility that both the flare and CME might contribute to the shock wave. But in the present case, the observational evidence supports flare shock scenario.

\section{Discussion}

 The present study provides a strong evidence of \textrm{EUV} wave being an MHD shock that excited the metric type II radio burst. The type II radio burst is observed by three \textrm{NRH} frequencies at 228, 173 and 150 MHz, which enables tracking the outward motion of harmonic band radio sources.  The \textrm{NRH} radio sources were located at the flank of the \textrm{EUV} wave observed in the \textrm{AIA 171 {\AA} and AIA 211 {\AA}} channels. There is a weak CME observed by \textrm{STEREO-B/SECCHI-COR1 and SOHO/LASCO} images.

The origin of coronal shocks that generate type II radio bursts is still under debate. Most of the observed type II radio bursts were accompanied by CMEs and are usually considered as shock driver. In the event presented herein, the \textrm{EUV} wave and the associated shock were seemingly triggered by the flare energy release. The radio sources located at the flank of the \textrm{EUV} wave evidences that the \textrm{EUV} wave act as the source of the radio emission. STEREO observed the CME in \textrm{STEREO-B/SECCHI-EUVI and SECCHI-COR1} images. Overplotting the \textrm{NRH} sources and the heights of eruptive structure in \textrm{STEREO-B/SECCHI-EUVI} on the AIA 171 {\AA} images show that radio sources are associated with the EUV wave and not with the CME leading edge. This indicates that the shock is probably ignited by the flare. Similarly, \cite{Wagner83} found in one event recorded by \textrm{Culgoora Radio Heliograph} data that the radio source was located well below the white-light transient before the event has reached the height of 2 $R_\odot$ and the shock was considerably faster than the CME. Thus they concluded that the type II radio burst is independent of the CME and a shock might be ignited by the associated flare. A similar conclusion was also reported by \citep{Vrsnak06} using \textrm{NRH} data.

 There are few case studies recently reported either without CME or with a slow CME maintains the debate on the origin of coronal shocks \citep{Magdalenic12, Su15, Eselevich17} whether flare or CME acts as a driver for their generation. We have to explore the context in detail in future studies. It is important to note that even now with the high resolution imaging observations from \textrm{SDO/AIA}, still we lack the observation of low coronal features for CMEs without bright fronts.

\section{Conclusions}

  Using combined data analysis, the detailed case study of a metric type II radio burst, that occurred on 06 March 2014 accompanied by an impulsive flare and a CME is reported. It is a classical radio event that contains type III and type II radio bursts. The type II radio burst is observed with both fundamental and harmonic structures in the frequency range between 235 $\--$ 130 MHz (120 $\--$ 60 MHz) in harmonic (fundamental) bands with a life time of 5 min between 09:26 UT $\--$ 09:31 UT. The estimated speed of the type II radio burst by employing $2\times$Saito electron density model is found to be $\sim$ 650 km s$^{-1}$.

  The type II radio burst is related to a coronal \textrm{EUV} wave clearly identified in \textrm{AIA 171, 211 and 193 {\AA}} images. \textrm{AIA 171 and 211 {\AA}} images were used to show the wave evolution. At \textrm{AIA 171 {\AA}} it has two components, one propagating parallel and other propagating perpendicular to the flare site. The parallel component gets reflected backward from one of the flare loops and the perpendicular component freely propagates outward. The wave is also observed in \textrm{AIA 211 {\AA}} images as a freely propagating outward moving structure. The estimated speed of the EUV wave at \textrm{AIA 171 {\AA}} is $\sim$ 820 km s$^{-1}$.

  The source position of the type II radio burst is observed at 228, 173 and 150 MHz \textrm{NRH} imaging frequencies. The NRH radio sources superposed on the \textrm{AIA 171 {\AA}} images reveals that the radio burst is excited at the outward moving \textrm{EUV} wave. The eruption is also accompained by a weak CME observed with a sky plane speed of $\sim$ 252 km s$^{-1}$ in SOHO/LASCO-C2 FOV and $\sim$ 280  km s$^{-1}$ in \textrm{STEREO-B/SECCHI-COR1 FOV}. The difference in speed between CME and type II burst implies that CME was probably not driving the shock. The h$\--$t plots of the type II burst and eruptive structures at different wavelengths show that the high-located EUV loops started to move outward, the EUV wave from behind created the type II burst and probably made the lower CME-materials to move faster. The flare energy release and the onset of EUV wave occurred simultaneously, as well as the speed between the EUV wave and type II burst lies closer to each other. Therefore the observed type II radio burst and the \textrm{EUV} wave are probably ignited by the flare.

\begin{acks}
This work was supported by the POB Anthropocene research program of Jagiellonian University, Krakow, Poland.
The author greatly acknowledge the various online data centers of
NOAA and NASA for providing the data. We express our thanks to the ORFEES,
LEARMONTH, NRH, SDO/AIA, STEREO/SECCHI, SOHO/LASCO teams for providing the data and LASCO CME
catalog used is generated and maintained by the Center for Solar
Physics and Space Weather, The Catholic University of America in
cooperation with the Naval Research Laboratory and NASA.
\end{acks}

\bibliographystyle{spr-mp-sola}

\bibliography{Vasanth}

    \begin{figure}[h]  
     \vspace{-0.01\textwidth}
      \centerline{\hspace{0.01\textwidth} \includegraphics[width=1.0\textwidth,angle=0]{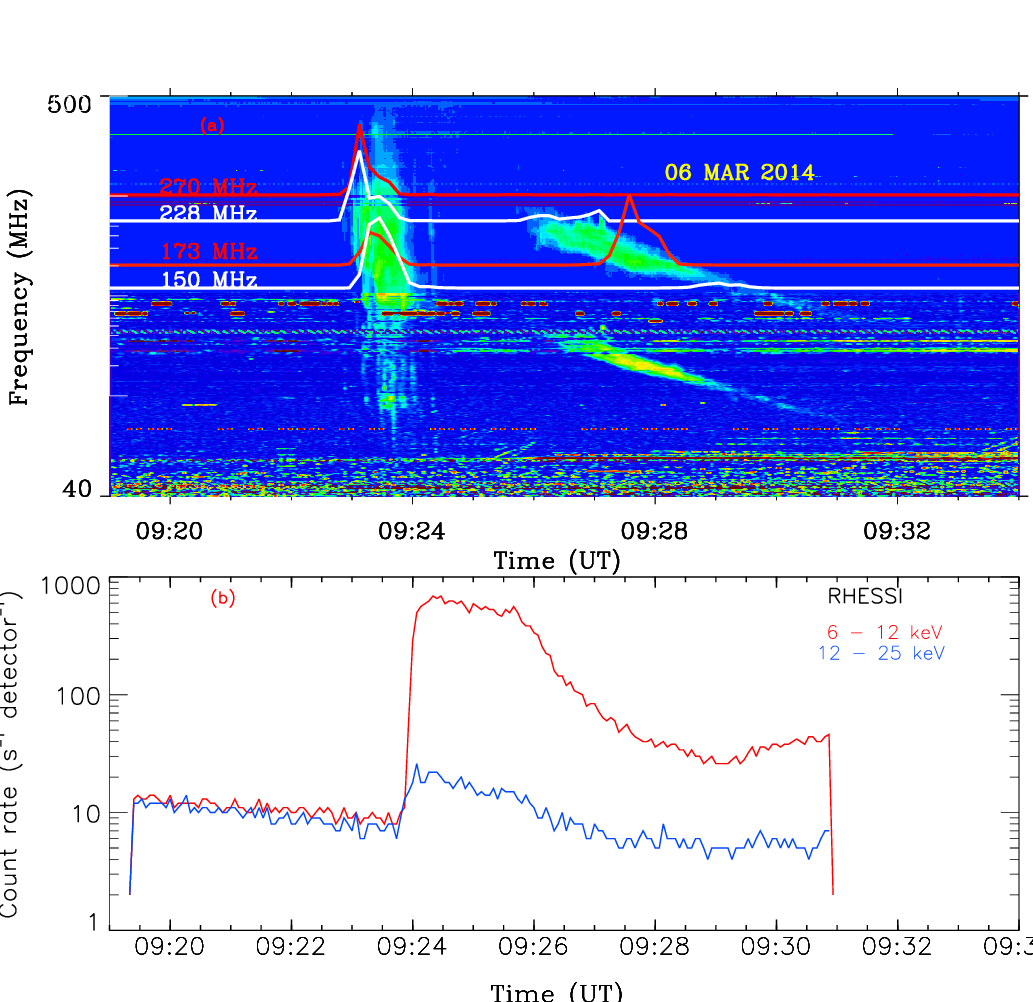}
                    }
    \vspace{0.03\textwidth}
        \caption{(a) Combined radio dynamic spectrum recorded by \textrm{ORFEES} (500 $\--$ 144 MHz) and \textrm{LEARMONTH} (180 $\--$ 25 MHz). Both the fundamental and harmonic components of the type II burst are clearly observed between 09:26 UT $\--$ 09:31 UT. The red lines (red labels) represent the total flux at 270 and 173 MHz, white lines (white labels) represent the total radio flux at 228 and 150 MHz obtained by \textrm{NRH} observations, (b) \textrm{RHESSI} X-ray profile in two enery bands at 6 $\--$ 12 and 12 $\--$ 25 keV.}
    \end{figure}

   \begin{figure}    
   \centerline{\includegraphics[width=1.0\textwidth,clip=]{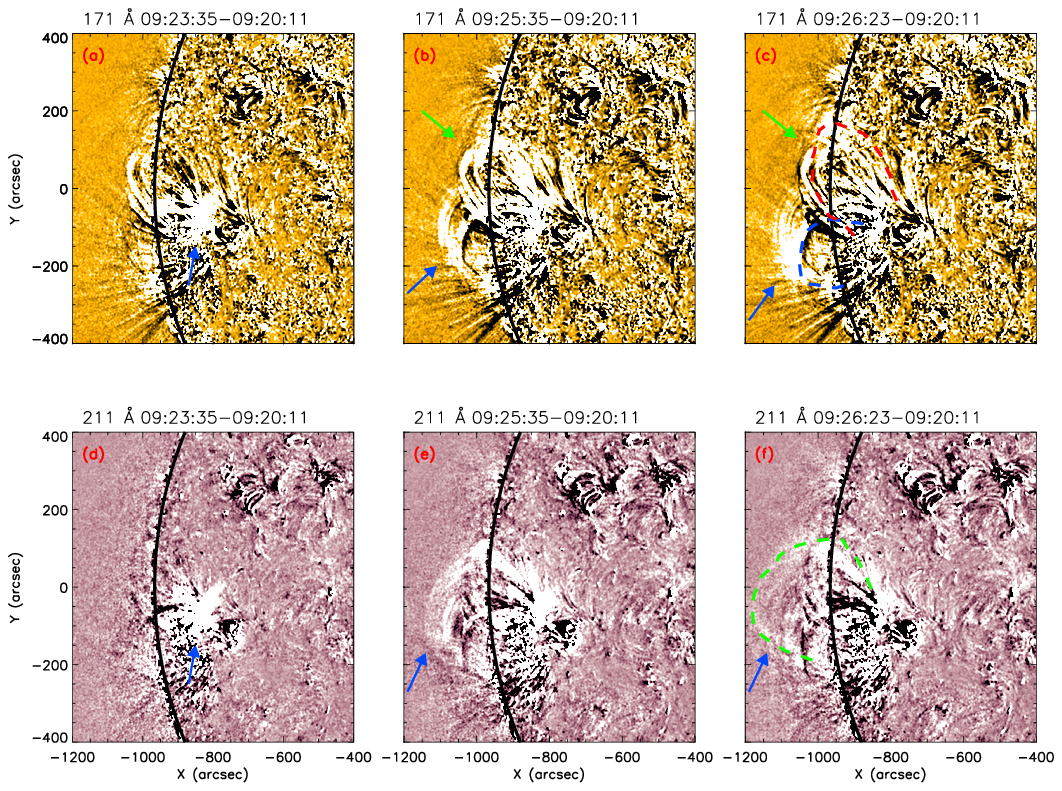}
              }
   \caption{(a - c) Base difference images at \textrm{AIA 171 {\AA}} showing the evolution of \textrm{EUV} wave components moving parallel (blue arrow) and perpendicular (green arrow) from the active region, (d - f) Base difference images at \textrm{AIA 211 {\AA}} showing the outward moving \textrm{EUV} wave structure indicated by green curve (blue arrow). An animation of the AIA images is available.}
    \label{F-appendix}
  \end{figure}

    \begin{figure}[h]   
        \vspace{-0.01\textwidth}
      \centerline{\hspace{-0.03\textwidth} \includegraphics[width=1.0\textwidth]{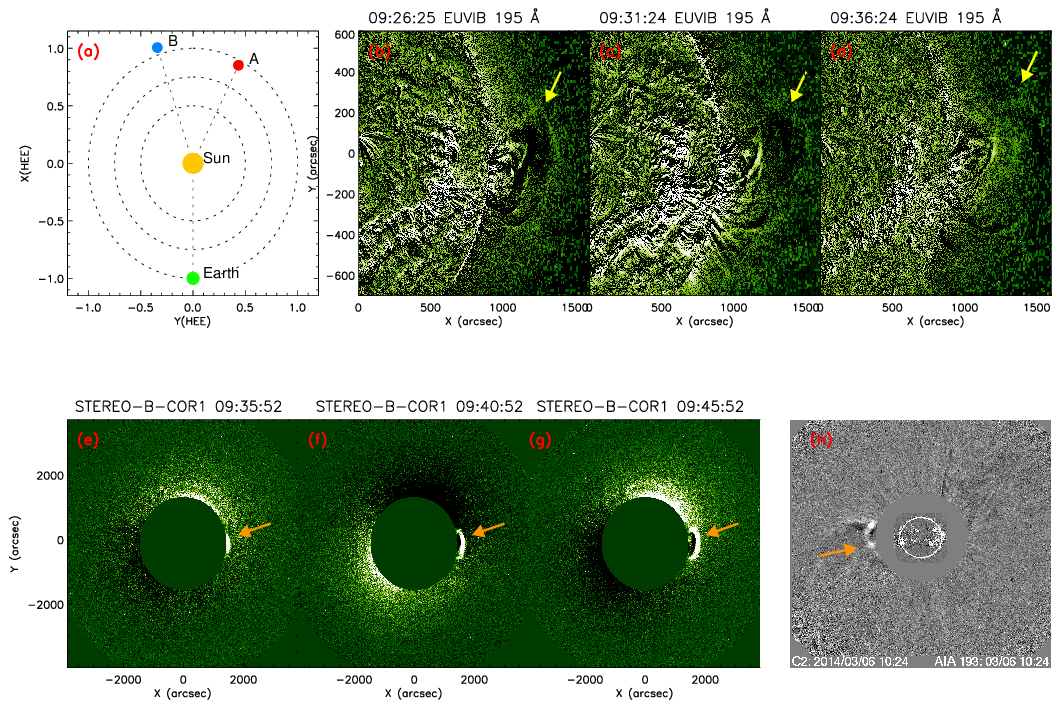}
                    }
        \vspace{-0.005\textwidth}
        \caption{(a) Spacecraft locations (A and B denotes the \textrm{STEREO} spacecraft positions, Sun denotes the Sun, Earth denotes the ground based observations (metric raio wavelengths),(b - d) snapshot of \textrm{STEREO-B/SECCHI-EUVI} running difference images at 195 {\AA} showing the eruptive structures (see yellow arrow), (e - h) The temporal and structural evolution of the CME by \textrm{STEREO-B/SECCHI-COR1} at 09:35 UT, 09:40 UT, 09:45 UT and \textrm{SOHO/LASCO-C2} at 10:24 UT (see orange arrow). An animation of the \textrm{STEREO-B/SECCHI-EUVI} running difference images is available.}
    \end{figure}

    \begin{figure}[h]   
        \vspace{-0.05\textwidth}
      \centerline{\hspace{0.01\textwidth} \includegraphics[width=1.0\textwidth]{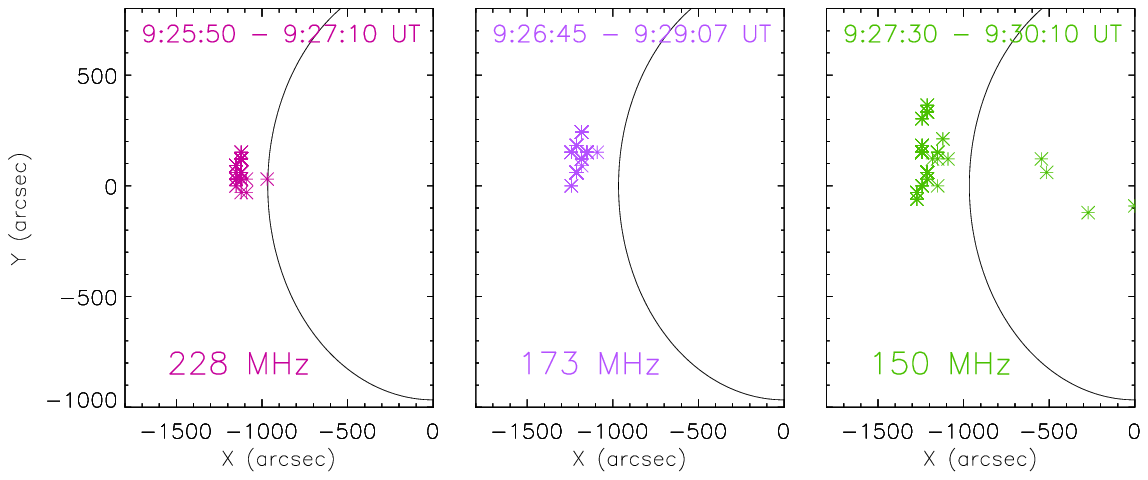}
                    }
    \vspace{0.08\textwidth}
    \caption{Temporal evolution of radio source centers from \textrm{NRH} images at different frequencies.An animation of the NRH radio imaging at different frequencies is available.}
   \end{figure}

    \begin{figure}[h]   
         \vspace{-0.05\textwidth}
        \centerline {\hspace*{-0.05\textwidth} \includegraphics[width=1.1\textwidth]{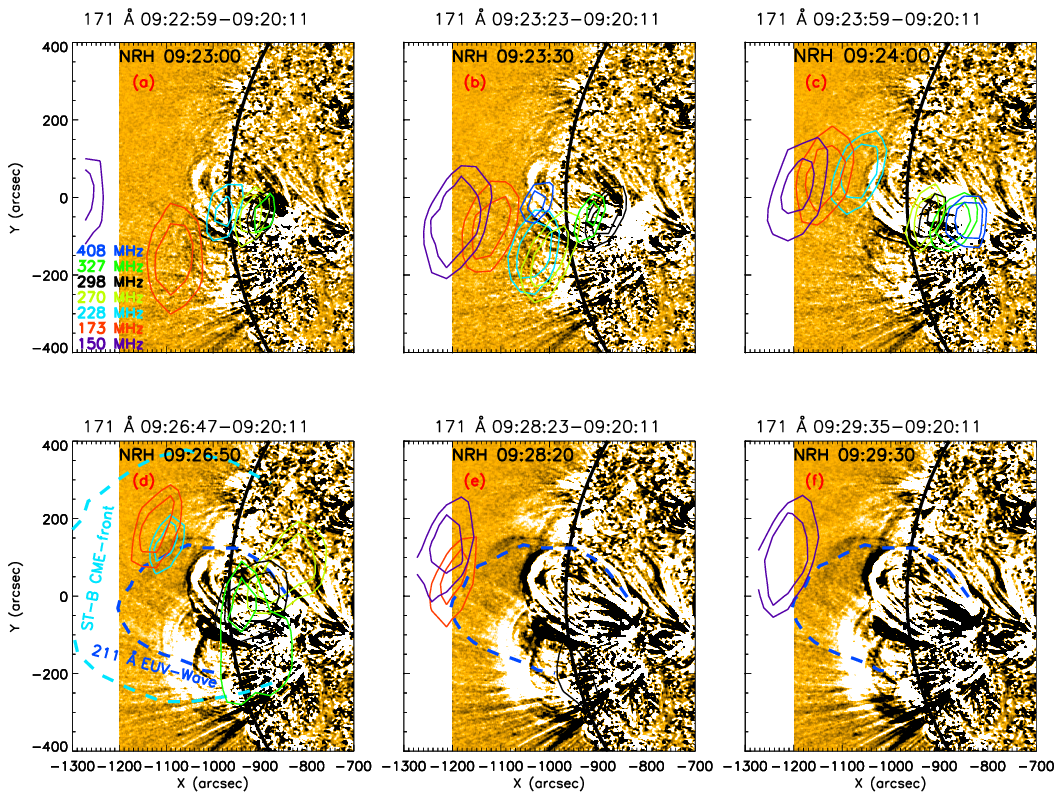}}
    \vspace{0.05\textwidth}
    \caption{Temporal evolution of \textrm{NRH} sources superposed on the closest-in-time of \textrm{AIA 171 {\AA}} images, the blue curve presents the \textrm{EUV} wave observed at \textrm{AIA 211 {\AA}} image and cyan color curve represents the CME-frontal structure observed by \textrm{STEREO-B/SECCHI-EUVI} image at closest-in-time. The upper row of images shows the type III burst locations to verify the frequency$\--$height dependence in plasma emission and the lower row the type II burst. The \textrm{NRH} data are represented by the 90 - 95 \% $T_{\rm B max}$ contour. An animation of this figure is available.}
    \end{figure}

   \begin{figure}[h]   
      \includegraphics[width=0.495\textwidth]{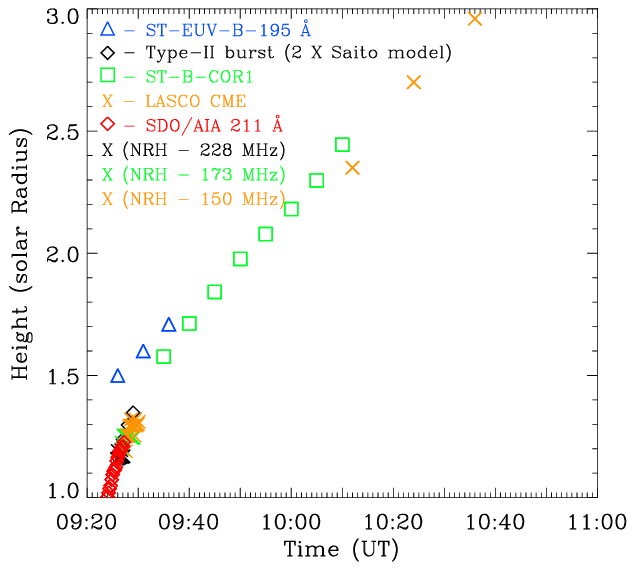}
      \includegraphics[width=0.495\textwidth]{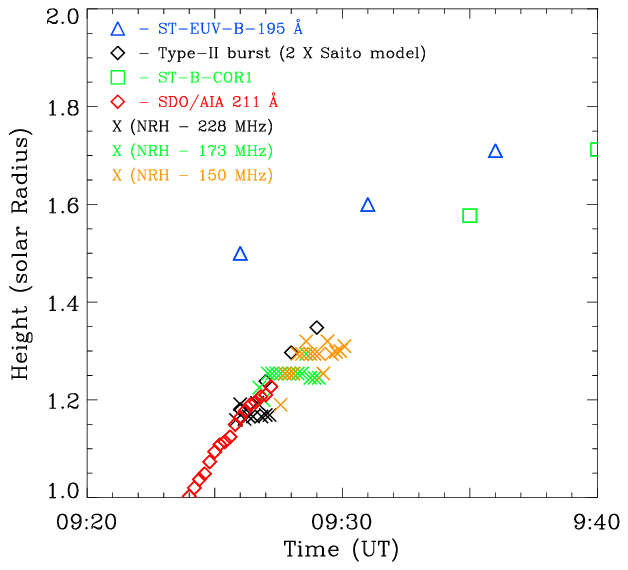}
    \caption{Kinematics of \textrm{EUV} wave and CME structures with metric type II shock signatures}.
   \end{figure}

\end{document}